\UseRawInputEncoding 
\documentclass[twocolumn,showpacs,preprintnumbers,amsmath,amssymb,superscriptaddress]{revtex4-1}

\usepackage{graphicx}
\usepackage{dcolumn}
\usepackage{bm}
\usepackage{color}
\usepackage{pifont}
\usepackage{natbib}
\usepackage{hyperref}
\hypersetup{
colorlinks = true,
urlcolor   = blue,
linkcolor  = blue,
citecolor  = blue
}

\begin{document}

\preprint{APS/PRB}
\title{Pressure driven magnetic order in  Sr$_{1-x}$Ca$_x$Co$_2$P$_2$}

\author{Ola Kenji Forslund}
 \email{okfo@kth.se}
\affiliation{Department of Applied Physics, KTH Royal Institute of Technology, SE-106 91 Stockholm, Sweden}
\author{Daniel~Andreica}
\affiliation{Faculty of Physics, Babes-Bolyai University, 400084 Cluj-Napoca, Romania​}
\author{Yasmine Sassa}
\affiliation{Department of Physics, Chalmers University of Technology, SE-41296 G\"oteborg, Sweden}
\author{Masaki~Imai}
\author{Chishiro~Michioka}
\author{Kazuyoshi~Yoshimura}
\affiliation{Department of Chemistry, Graduate School of Science, Kyoto University, Kyoto 606-8502 Japan}
\author{Zurab Guguchia}
\author{Zurab Shermadini}
\author{Rustem Khasanov}
\affiliation{Laboratory for Muon Spin Spectroscopy, Paul Scherrer Institute, CH-5232 Villigen PSI, Switzerland}
\author{Jun Sugiyama}
\affiliation{Neutron Science and Technology Center, Comprehensive Research Organization for Science and Society (CROSS), Tokai, Ibaraki 319-1106, Japan}
\author{Martin M\aa nsson}
 \email{condmat@kth.se}
\affiliation{Department of Applied Physics, KTH Royal Institute of Technology, SE-106 91 Stockholm, Sweden}

\date{\today}

\begin{abstract}
The magnetic phase diagram of Sr$_{1-x}$Ca$_x$Co$_2$P$_2$ as a function of hydrostatic pressure and temperature is investigated by means of high pressure muon spin rotation, relaxation and resonance ($\mu^+$SR). The weak pressure dependence for the $x\neq1$ compounds suggests that the rich phase diagram of Sr$_{1-x}$Ca$_x$Co$_2$P$_2$ as a function of $x$ at ambient pressure may not only be attributed to solely chemical pressure effects. The $x=1$ compound on the other hand reveals a high pressure dependence, where the long range magnetic order is fully suppressed at $p_{\rm c2}\approx9.8$~kbar, which seem to be a first order transition. In addition, an intermediate phase consisting of dilute ferromagnetic islands (FMI) is formed above $p_{\rm c1}\approx8$~kbar where they co-exist with a magnetically disordered state. Moreover, such FMI phase seems to consist of an high- (FMI-\textcircled{\small{1}}) and low-temperature (FMI-\textcircled{\small{2}}) region, respectively, separated by a phase boundary at $T_{\rm i}\approx20$~K.

\end{abstract}



\maketitle

\section{\label{sec:Intro}Introduction}
\begin{figure}[ht]
  \begin{center}
    \includegraphics[keepaspectratio=true,width=85mm]
    {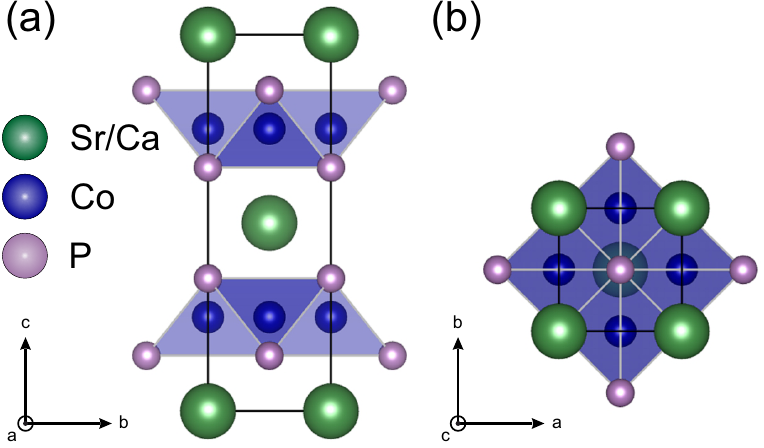}
  \end{center}
  \caption{The crystal structure of Sr$_{1-x}$Ca$_{x}$Co$_{2}$P$_{2}$ shown in two planes: (a) the cb plane and (b) the ab plane. The green spheres represents Sr/Ca atoms while the blue and pink spheres represents Co and P atoms, respectively. The structure is stabilised in Tetragonal symmetry with space group I4/mmm (\#139).}
  \label{fig:Crystal}
\end{figure}

The ThCr$_2$Si$_2$ layered structure type family of compounds typically exhibits ground states ranging from superconductivity to long range magnetic order \cite{Johrendt1997, Reehuis1998, Hoffmann2002, Rotter2008, Tan2017}. The ground states are determined by competition between magnetism and superconductivity, as in CaFe$_2$As$_2$ or Ba$_{1-x}$K$_x$Fe$_2$As$_2$ \cite{Rotter2008_2, Torikachvili2008}. In this family, the $AM_2X_2$ structure type with a metal $A$, transition metal $M$, and metalloid $X$ atoms, are made up of edge-share $T´MX_4$ tetrahedra layers (Fig.~\ref{fig:Crystal}). The delicate ground state is mostly dependent on the inter-layer $X$-$X$ bonding distance across the intermediate $A$ sheets. For several $AM_{2}$P$_{2}$ ($A$ = Ca, Sr, and Ba, and $M$ = Fe, Co, and Ni) materials, the phase transitions are related to subtle structural changes present in the crystals. In particular, these families have a tendency to transform from uncollapsed tetragonal (ucT) to collapsed tearagonal (cT) structure. This is driven by the X-X bonding acting between the $M_2X_2$ layers, for which strong enough bonds pull the layers closer and induces a lattice relaxation. \cite{Reehuis1990,Kreyssig2008,Shuang2009}

In the case of Sr$_{1-x}$Ca$_{x}$Co$_{2}$P$_{2}$, the crystals transforms from ucT to cT when the chemical composition changes from $x=0$ to $x=1$. A Curie-Weiss behaviour is observed in all compounds at high temperature, but the high temperature fluctuations changes from antiferromagnetic to ferromagentic type around $x\sim0.5$ \cite{Shuang2009}. Such change seem to be correlated to the finally ground state as it is transformed from paramagnetic ($x<0.45$) to antiferromagnteic (AF) at $x\approx0.5$ \cite{Sugiyama2015}. In fact, bulk magnetisation measured as a function of $x$ show a clear correlation between the detailed crystalline structure and the magnetic properties \cite{Shuang2009}.

Previous study based on muon spin rotation, relaxation and resonance ($\mu^+$SR) \cite{Sugiyama2015} have indicated Pauli-paramagnetic phases for $x<0.45$ at temperatures as low as 1.8 K, consistent with De Haas-van Alphen measurement \cite{Teruya2014}. Short-range AF ordered phases appears for $0.48\leq x\leq 0.75$, which stabilizes into a long-range AF ordered phase for $x>0.75$. The formation of magnetically ordered phases were shown to have strong correlations with the nearest neighboring Co distance ($d_{\rm Co-Co}$) within the Co$_{2}$P$_{2}$ planes, implying the importance of also the inter-plane interaction for the formation of long range order. The Co-Co distance decreases only moderately with $x$ up until $\sim0.5$, at which point $d_{\rm Co-Co}$ experience an abrupt decrease until 0.9 where it finally levels out to an almost constant value \cite{Sugiyama2015}.

It is thus interesting to continue the investigation by further decreasing the distance $d_{\rm Co-Co}$ through the application of a hydrostatic pressure. Such study was performed on CaCo$_{2}$P$_{2}$ single crystals using resistivity \citep{Baumbach2014}. These measurements suggested that the AF order is destabilises with pressure, hinting towards a quantum critical point (QCP). Moreover, the signature of a second unknown phase was also indicated in the pressurised state, but the details of such phase remain unknown.

In order to confirm and further comprehend the high pressured states, we have conducted pressure dependent $\mu^{+}$SR measurements on the series of Sr$_{1-x}$Ca$_{x}$Co$_{2}$P$_{2}$ powder samples, including $x=1$, 0.9, 0.8 and 0.7. Given that $\mu^{+}$SR is highly sensitive to magnetic fields and to magnetic volume fraction, any subtle magnetic transition can be detected and characterized in detail. The current study clearly reveal that the AF order is suppressed with pressure, consistent with macroscopic measurements. As we shall show, this transition does however seem to be first order type, thereby ruling out the possibility of a QCP. Moreover, the sample exhibits a co-existence of magnetic order and disorder at higher pressures just below the critical pressure.

\section{\label{sec:exp}Experimental Setup}
Sr$_{1-x}$Ca$_x$Co$_2$P$_2$ polycrystalline samples were synthesised in a two step reaction from the base elements; P, Sr, Ca, and Co. At first, Sr, Ca, Co were individually put together with P in an evacuated quartz tube to facilitate a solid state reaction at 800$^\circ$C and 700$^\circ$C to produce SrP, CaP and Co$_2$P. Sr$_{1-x}$Ca$_x$Co$_2$P$_2$ was then synthesized via a solid-state reaction between SrP, CaP, and Co$_2$P at 1000$^\circ$C for 20 hours in Ar atmosphere. Detailed information about the synthesis protocol is found in Ref.~\onlinecite{Imai2014}.

The $\mu^{+}$SR measurements were performed at the GPD instrument on the $\mu$E1 beamline at the S$\mu$S muon source of Paul Scherrer Institute (PSI), Switzerland. Hydrostatic pressures up to 23 kbar were achieved by using a piston cylinder cell made of MP35 and CuBe alloys, respectively. Temperatures down to $T=2$ K were achieved using a $^{4}$He flow cryostat. Three pressed pellets of the powder samples were stacked (5.9 mm diameter and 13 mm total height) for each measurement and inserted into the pressure cell. Daphne oil was used as the pressure medium in order to achieve the hydrostatic pressure. The pressure of the sample at low temperatures was accurately determined via AC susceptibility measurements of the superconducting transition temperature for a small indium wire located at the bottom of (inside) the pressure cell. \cite{PSI_GPD1, PSI_GPD2} Finally, the free analysis software \texttt{musrfit} \cite{musrfit} was used to analyzed the $\mu^{+}$SR data. 

\section{\label{sec:results}Results}
Series of $\mu^+$SR measurements on Sr$_{1-x}$Ca$_x$Co$_2$P$_2$ are presented as a function of pressure, temperature and chemical composition ($x=$~1, 0.9, 0.8 and 0.7). In particular, measurements in weak transverse field (TF) configuration are used in other to estimate the phase boundaries, while zero field (ZF) measurements are employed to deduced the detailed magnetic ground state. Transverse refers to the direction of the externally applied field, which is perpendicular with respect to the initial muon spin polarisation.

\begin{figure*}[ht]
  \begin{center}
    \includegraphics[keepaspectratio=true,width=\textwidth]
    {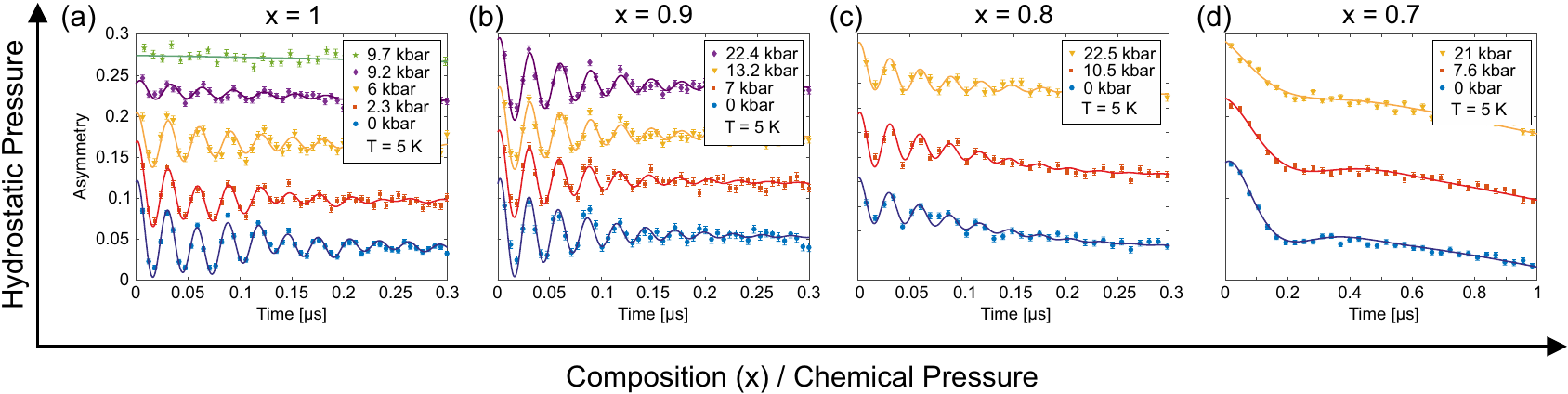}
  \end{center}
  \caption{Zero field time spectra collected at selected pressures at $T=5$~K for the Sr$_{1-x}$Ca$_x$Co$_2$P$_2$ compounds: (a) $x=1$, (b) $x = 0.9$, (c) $x=0.8$ and (d) $x=0.7$. The solid lines represents fits using Eq.~(\ref{eq:ZF}). Each spectra have been shifted vertically for clarity of display.
  }
  \label{fig:ZFSpec}
\end{figure*}

\subsection{\label{sec:results}Zero field}
Figure~\ref{fig:ZFSpec} displays the collected ZF $\mu^+$SR time spectra at $T=5$~K for the compositions $x=1$, 0.9, 0.8 and 0.7 as a function of pressure. The time spectra for the compounds at 0~kbar are fully consistent with previously published ambient pressure results \cite{Sugiyama2015}. A strong pressure dependence is observed for the $x=1$ compound. Indeed, several oscillations are present in the time spectra for $p\leq9.2$~kbar, where the number of frequencies as well as the amplitude of the oscillations decreases with the applied pressure. Such oscillations are fully suppressed for $p_{\rm c2}\geq9.7$~kbar. A much more moderate pressure dependence is observed for the other compositions ($x=$~0.9, 0.8 and 0.7). Consequently, it is initially clear that the ZF time spectra has a strong $x$ dependence, while the most evident hydrostatic pressure effect seem to be limited to $x=1$ compound. In order to more systematically characterize the detailed changes across $x$ and $p$, the time spectra were fitted using a combination of several exponentially relaxing oscillations, an exponential and a static Gaussian Kubo-Toyabe (SGKT):

\begin{eqnarray}
 A_0 \, P_{\rm ZF}(t) &=&
\sum^{n}_{i} A^{\rm AF}_{i} \cos(f^{\rm AF}_i2\pi t+\phi^{\rm AF}_i)e^{-\lambda^{\rm AF}_i t}
\cr\cr
&+&A_{\rm tail}e^{-\lambda_{\rm tail}t}+A_{\rm PC}G(t,\Delta_{\rm PC})e^{-\lambda_{\rm PC}t}
\cr\cr
&+&A_{\rm PM}e^{-\lambda_{\rm PM}t},
\label{eq:ZF}
\end{eqnarray}

where $A_{0}$ is the initial asymmetry, determined by the instrument, and $P_{\rm ZF}$ is the muon spin polarisation function in ZF configuration. In detail, the number of oscillations, $n=3$ and $n=2$, where used for the compositions $x=1$ and $x=0.8$, respectively, while $n=1$ was used for both $x=0.9$ and $x=0.7$. $A^{\rm AF}_i$, $f^{\rm AF}_i$, $\phi^{\rm AF}_i$ and $\lambda^{\rm AF}_i$ are the asymmetry, precession frequency, phase and relaxation rate for the internal field component that is perpendicular with respect to the initial muon spin polarisation. $A^{\rm tail}$ and $\lambda^{\rm tail}$ on the other hand are the asymmetry and relaxation rate of the so called tail component, $i.e.$ the internal field component that is parallel to the initial muon spin polarisation. In a perfect powder that is magnetically ordered, 2/3 of the internal field components are expected to be perpendicular while 1/3 of the internal field components are parallel (due to the so-called 'powder average'). $A_{\rm PM}$, and $\lambda_{\rm PM}$ are accounting for the new high pressure paramagnetic (PM) phase of the $x=1$ compound that is not magnetically ordered [Fig.~\ref{fig:ZFSpec}(a)]. $A_{\rm PC}$, $\Delta_{\rm PC}$ and $\lambda_{\rm PC}$ on the other hand are the asymmetry, the field distribution width and the corresponding exponential relaxation rate of the static Gaussian KT, represented by $G(t, \Delta)$, of the pressure cell (PC). Here, $G(t, \Delta)$ originate from isotropically distributed magnetic moments while the exponential accounts for additional relaxation present on top of it. Such description holds for when the internal field is composed of two separate yet independent magnetic field origins. In such case, the Fourier transform of the convolution of each field distribution is the product of each polarisation functions. In other words, the KT accounts for isotropically distributed nuclear moments while the exponential accounts for additional relaxation posed by highly fluctuating electronic moments.
  
\begin{figure}[ht]
  \begin{center}
    \includegraphics[keepaspectratio=true,width=75mm]
    {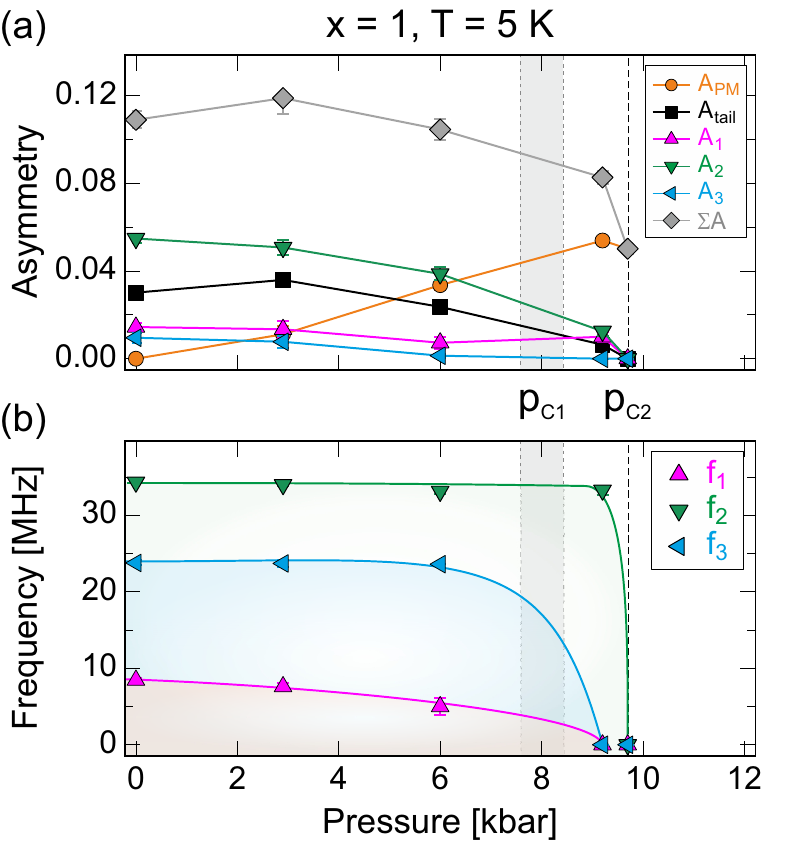}
  \end{center}
  \caption{Pressure dependent fit parameters, collected at $T=5$~K, of the $x=1$ compound obtained through Eq.~(\ref{eq:ZF}). For reference, $\sum A=\sum A^{\rm AF}_i+A_{\rm tail}+A_{\rm PM}$ has been computed and is included as well. The solid lines are guide to the eye while the vertical shaded area and dashed line indicate the critical pressures, $p_{\rm c1}\approx8$~kbar and $p_{\rm c2}'\approx9.5$~kbar.
  }
  \label{fig:ZFPara_1}
\end{figure}

The constraint $A_{\rm tail}=\frac{1}{2}\sum A^{\rm AF}_i$ (i.e. 1/3 vs. 2/3) was set for the fitting procedures using Eq.~(\ref{eq:ZF}), in order to separate the various contributions present for $p>0$. Moreover, the compounds are known exhibit commensurate magnetic order. Indeed, $\phi_1=-20.0(2.9)^\circ$, $\phi_2=-10.6 (10.4)^\circ$ and $\phi_3= -28.7(16.6)^\circ$ are obtained for the compound $x=1$ at $p=0$, consistent with a commensurate order. Therefore, a common phase was set, $i.e.$ $\phi_i=\phi$, across all oscillations for all measured pressures. Using the described fitting procedure, the pressure dependencies of the fit parameters for the $x=1$ compound are shown in  Fig.~\ref{fig:ZFPara_1}. At ambient pressure $A_{\rm tail}\simeq\frac{1}{2}\sum A^{\rm AF}_i$ is obtained, suggesting that the constraint set for higher pressures is valid. For higher pressures, the asymmetries $A^{\rm AF}_i$ decreases gradually as the $A^{\rm PM}$ component increases.

\begin{figure}[ht]
  \begin{center}
    \includegraphics[keepaspectratio=true,width=75mm]
    {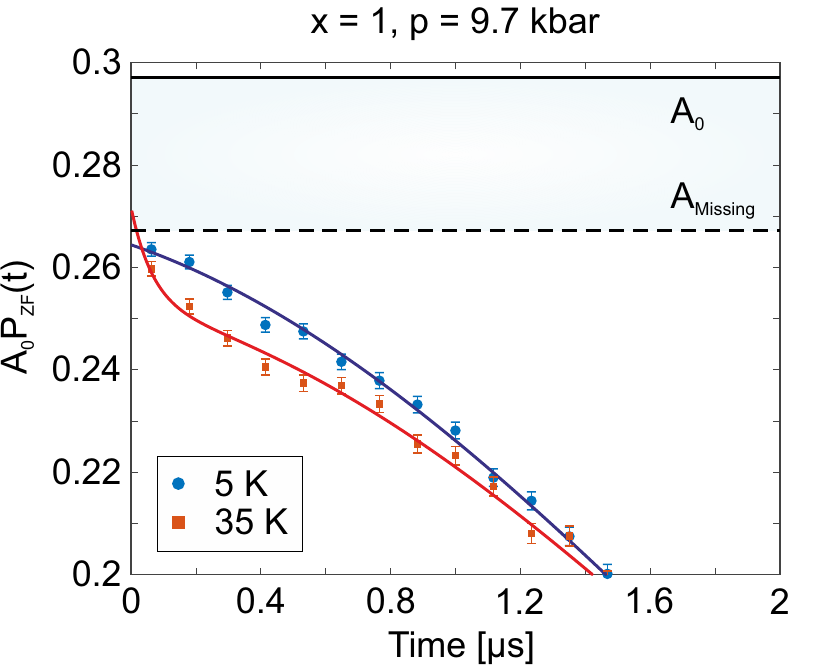}
  \end{center}
  \caption{Zero field time spectra collected at $T=5$ and 35~K at $p=9.7$~kbar for the $x=1$ compound. The solid lines represents fits using Eq.~\ref{eq:ZF} with $n=1$ but with $f_1=0$~MHz for $T=35$~K and $n=0$ for $T=5$~K. The horizontal solid black line is the estimated $A_0$ while the dashed black horizontal line indicate the amount of asymmetry missing.
  }
  \label{fig:ZFSpec_1}
\end{figure}

It should be noted that the total sample asymmetry ($\sum A=\sum A^{\rm AF}_i+A_{\rm tail}+A_{\rm PM}$) show a fairly constant behaviour up to about 6~kbar. At higher pressures, a sudden decreases is observed. The origin of this behaviour is highlighted in Fig.~\ref{fig:ZFSpec_1}. As clearly seen, a significant missing fraction presents itself at higher pressures. Since a missing fraction cannot be fitted (since it lies outside the accessible time frame of the $\mu^+$SR instrument), the total sample asymmetry ($\sum A$) exhibit a decrease as a function of pressure. The origin of this missing fraction is discussed below. As mentioned, $A^{\rm PM}$ represents the volumic fraction of non-magnetically ordered state, $e.g.$ paramagnetic or spin liquid. Such high pressure state is discussed in Sec.~\ref{sec:discussion}.

The pressure dependence of the precession frequencies for the $x=1$ compound are shown in Fig.~\ref{fig:ZFPara_1}(b). Overall, the pressure seem to have a weak effect on the value of the frequencies themselves (i.e. the local internal magnetic field of the sample). The main frequency, $f_2$, maintains more or less the value at ambient pressure ($\sim34$~MHz) up to 9.2~kbar ($\sim33$~MHz). We may initially define the pressure point at which all precession frequencies are absent as the critical pressure, $p_{\rm c2}'\approx9.5$~kbar. We will further refine such critical point based on the TF measurement presented below. It is noted that the 24~MHz ($f_3$) frequency was not present in the previous ambient pressure study \cite{Sugiyama2015}. Most likely, the $x=1$ sample of the present study is both much larger and of higher quality, resulting in that even the minor frequency can be distinguished. That being said, the inclusion of such frequency does not change/alter the interpretation of this or the previous study. 

Further, it is evident that the $f_1$ and $f_3$ components drop to 0~MHz already at $p_{\rm c1}\approx8$~kbar, i.e. prior to the vanishing of the $f_2$ component at $p_{\rm c2}'\approx9.5$~kbar. It is however clear the asymmetry ($A^{\rm AF}_1$) still poses non-zero values. In other words, the polarisation is a fast exponential instead of an oscillation. Such behaviour suggests a widening of the internal field distribution width, in which the oscillation becomes highly damped. Such behaviour would be consistent with a spin reorientation, structural transition ($i.e.$ muon sites changes) and/or it could reflect a difficulty in fitting small asymmetries. However, we will later show (below) that this effect is indeed a true sample effect and not related to a fitting problem.


While it is not shown, the relaxation rates, $\lambda^{\rm AF}_i$, show a weak pressure dependence. Roughly put, a value of $\lambda^{\rm AF}_i=9~\mu$s$^{-1}$ is obtained for all oscillation across all pressures. $\lambda_{\rm tail}$ on the other hand exhibits low or values close to 0 across all pressures, suggesting a static magnetic ground state. $\lambda_{\rm PM}$ on the other hand increases relatively for $p>2.3$~kbar, suggesting that the high pressure phase is dynamic in origin. 

The ZF time spectra collected at 5 and 35~K for the $x=1$ at 9.7~kbar are shown in Fig.~\ref{fig:ZFSpec_1}. The 5~K time spectra is identical to the one shown in Fig.~\ref{fig:ZFSpec}(a). While the 5~K time spectrum show no significant signature, the 35~K time spectrum manifests an fast relaxing exponential. Perhaps a dip/minimum is present in Fig.~\ref{fig:ZFSpec_1} around $0.5~\mu$s, suggesting that the polarisation function is more of an oscillation rather than an exponential. However, a fit with a cosine function yields unreasonably high asymmetry values. Regardless, the fraction of this fast exponential decreases with decreasing temperature. Such behaviour is the origin behind the complex temperature dependence of $A_{\rm TF}$ at lower temperatures (Fig. \ref{fig:TFPara_1}(a)). In fact, resistivity measurements under pressure indicated a sudden change in the derivative of the resistivity around this temperature and pressure \citep{Baumbach2014} and is discussed in Sec. \ref{sec:discussion}. 

As already noted, a significant fraction is missing, $i.e.$ the total asymmetry does not add up to values close to $A_0$. Typically, missing fractions are associated with muonium formations \cite{Boekema1986, Cox2006, Forslund2018} or by a presence of quasi-static wide field distribution. Since the $x=1$ compound is metallic even under pressure, muonium formation is unlikely. Instead, the missing fraction originates from a wide internal field distribution yielding oscillations and fluctuations outside the time window of $\mu^+$SR. Such a scenario is consistent with having muon sites close to CoP$_2$ tetrahedra layers giving a wide field distributions due to slowly fluctuating Co $d$-moments. Similar missing fraction effect was observed e.g. in the 2D AF magnet NaNiO$_2$ \cite{Forslund2020}, which was shown to originate from quasi static wide field distributions at the muon site close to the Ni-O octahedra. In other words, the suppression of asymmetries [Fig.~\ref{fig:ZFPara_1}(a)] is consistent with changes in the magnetic characteristics at high pressure, giving the wider field distributions at the muon site ($i.e.$ the magnetic environment around the muon site changes). Given the crystal flexibility of the $x=1$ compound, it could also be that additional energetically favorable crystalline muon sites becomes available under higher pressure instead ($i.e.$ the crystalline muon site changes). Such assertion is ideally confirmed by a combination of high pressure XRD and DFT calculations. Regardless of the driver for the missing fraction, it is clear that the origin is related to magnetism of the sample since the missing fraction is in fact absent in the $p=17.2$~kbar measurement.

\begin{figure}[ht]
  \begin{center}
    \includegraphics[keepaspectratio=true,width=75mm]
    {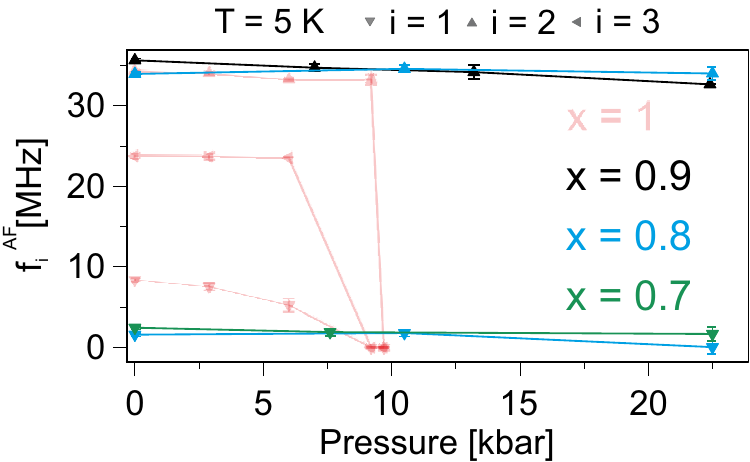}
  \end{center}
  \caption{Pressure dependent precession frequencies, collected at $T=5$~K, of Sr$_{1-x}$Ca$_x$Co$_2$P$_2$ compounds obtained through Eq.~(\ref{eq:ZF}): $x=1$ (light pink), 0.9 (black), 0.8 (blue) and 0.7 (green). The solid lines are guide to the eye. }
  \label{fig:ZFPara_Fre}
\end{figure}

For the sake of completeness, the pressure dependent precession frequencies for all measured $x$ is presented in Fig.~\ref{fig:ZFPara_Fre}. The frequencies of the $x=1$ compound are the same as presented in Fig.~\ref{fig:ZFPara_1}(b) (and are therefore made partly transparent). The main frequency of about 34 MHz is present for $x\geq0.8$. Such frequency is a consequence of the long range order present at ambient pressure for $x\geq0.8$~\cite{Sugiyama2015}. At $x=0.8$ however, an additional frequency appears around 2~MHz. Such frequency is concluded to be associated with a short-range order based on the fact that the same frequency but as a highly damped oscillation is present for the $x=0.7$ sample \cite{Sugiyama2015}. Perhaps $x=0.8$ exhibits a co-existence of short and long range order, similar to in the isostructural compound LCo$_2$P$_2$ \cite{Forslund2021_La}. Regardless though, only weak shifts in the precession frequencies are observed as a function of pressure for $x<1$. Similarly, the asymmetries and depolarisation rates show only weak pressure dependence, as already hinted directly from Fig.~\ref{fig:ZFSpec}(b-d).

\subsection{\label{sec:results}Transverse field}
\begin{figure}[ht]
  \begin{center}
    \includegraphics[keepaspectratio=true,width=75mm]
    {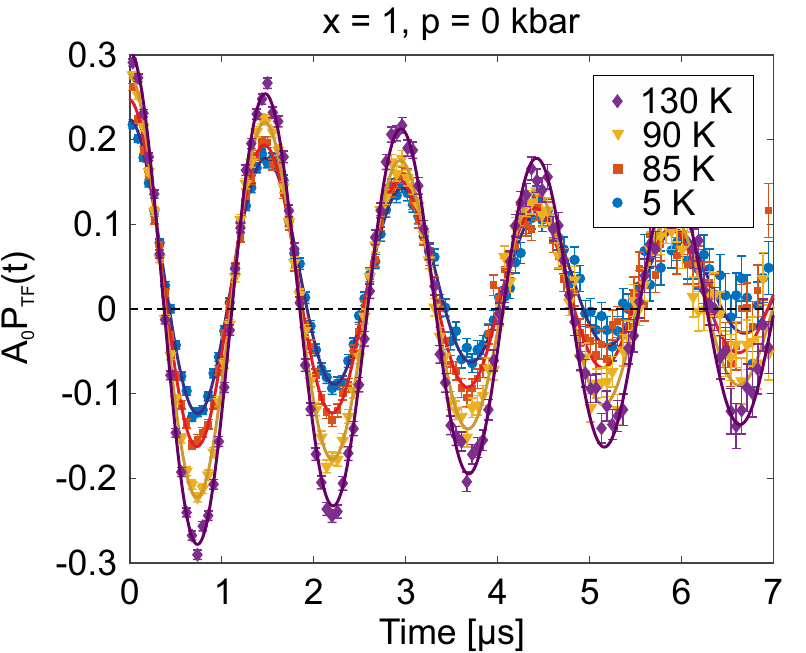}
  \end{center}
  \caption{Transverse field (TF~=~50~Oe) $\mu^+$SR time spectra collected for CaCo$_2$P$_2$ ($x=1$) at $p=0$~kbar for selected temperatures. The solid lines represents fits with Eq.~(\ref{eq:TF}).  
  }
  \label{fig:TFSpec_1}
\end{figure}

In order to gain a more detailed insights of the temperature dependent behaviour, the series of compounds were also studied under TF configuration for selected pressures. Figure~\ref{fig:TFSpec_1} shows the collected TF ($\sim50$~Oe) $\mu^+$SR time spectra from the $x=1$ compound at $p=0$ for selected temperatures. Regardless of temperature, a single distinct oscillation of about 0.7~MHz is observed, corresponding to the externally applied field of 50~Oe. A strong temperature dependence in the amplitude is seen, reflecting the formation of static internal magnetic fields. Moreover, the time spectra exhibits a positive shift in asymmetry at lower temperatures. Therefore, the TF time spectra were fitted using a combination of one exponentially relaxing oscillation together an non oscillating exponential:

\begin{eqnarray}
 A_0 \, P_{\rm TF}(t) &=&
A_{\rm TF} \cos(f_{\rm TF}2\pi t+\phi_{\rm TF})e^{-\lambda_{\rm TF}t}
\cr\cr
&+& A_{\rm S}e^{-\lambda_{\rm S} t},
\label{eq:TF}
\end{eqnarray}

where $A_{0}$ is the initial asymmetry determined by the instrument and $P_{\rm TF}$ is the muon spin polarisation function in TF configuration. $A_{\rm TF}$, $f_{\rm TF}$, $\phi_{\rm TF}$ and $\lambda_{\rm TF}$ are the asymmetry, frequency, phase and depolarisation rate resulting from the applied TF, while $A_{\rm S}$ and $\lambda_{\rm S}$ are the asymmetry and the relaxation rate resulting from internal magnetic field components, that is mostly parallel to the initial muon spin. The perpendicular internal magnetic field components are excluded from Eq.~(\ref{eq:TF}), since such contribution are usually difficult to model in TF configuration. Excluding such contribution does not affect the fit results nor the interpretation of the data. 

\begin{figure}[ht]
  \begin{center}
    \includegraphics[keepaspectratio=true,width=83mm]
    {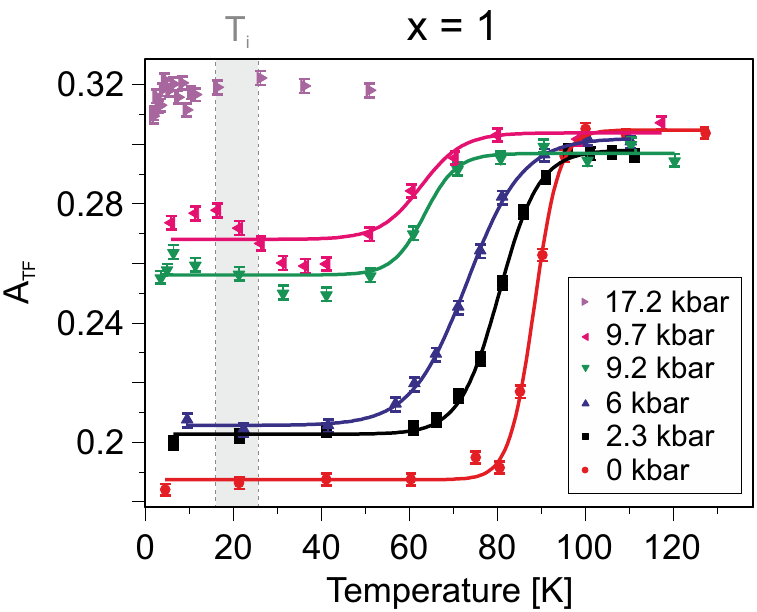}
  \end{center}
  \caption{Temperature and pressure dependence of $A_{\rm TF}$ for the $x=1$ compound, obtained using Eq.~(\ref{eq:TF}). The solid lines are fits using the sigmoidal function, while the vertical shaded area indicate an approximative region for the transition temperature $T_{\rm i}\approx20$~K (see text and Fig.~\ref{fig:Phase} for further details). 
  }
  \label{fig:TFPara_1}
\end{figure}

The temperature dependencies of the obtained TF asymmetry using Eq.~(\ref{eq:TF}) are displayed in Fig. \ref{fig:TFPara_1}(a) for the $x=1$ compound. $A_{\rm TF}$ has a temperature dependence expected for a magnetically ordered sample. At low temperatures $A_{\rm TF}$ experiences a temperature independent behaviour. As the temperature increases, an increase in $A_{\rm TF}$ is observed and full asymmetry ($A_0$) is recovered. Ideally, the normalized $A_{\rm TF}$/$A_0$ should corresponds to the paramagnetic volume fraction of the system. In this case, it is in fact the non-magnetically ordered volume fraction of the sample together with the fraction of muons stopping inside the pressure cell. In other words, the increase of $A_{\rm TF}$ corresponds to the transition temperature ($T^{\rm TF}_{\rm N}$) in which the sample changes from a magnetically ordered to disordered state. An accurate value of $T^{\rm TF}_{\rm N}$ is obtained by employing sigmoidal fit function as a function of temperature for each measured pressure. The pressure dependent values [i.e. ($T^{\rm TF}_{\rm N}(p,x)$] are then utilized to contruct the detailed phase diagram presented in Fig.~\ref{fig:Phase}. Intriguingly, $A_{\rm TF}$ show a complex temperature dependence below $T^{\rm TF}_{\rm N}$ for the measurements performed at $p=9.2$ and 9.7~kbar. In fact some of the TF asymmetry ($A_{\rm TF}$) seems to be recovered around $T_{\rm i}\approx20$~K (Fig. \ref{fig:TFPara_1}). Such complexity is reflecting the fact that an initial faster relaxation is present at higher temperature in ZF (Fig. \ref{fig:ZFSpec_1}). That $A_{\rm TF}$ still exhibits a drop at $T_{\rm N}$ is evidence of static internal field formation, which in ZF configuration resulted into a missing fraction as described above. As mentioned, this fraction stems from muons experiencing very broad field distribution width. At highest pressure on the other hand, no significant temperature dependence is observed, suggesting the absence of magnetic order at lower temperatures. A similar complicated $ A_{\rm TF}(T)$ behaviour is not observed in any of the other ($x\neq1$) compounds. 

\subsection{\label{sec:results}Phase diagram}
\begin{figure*}[ht]
  \begin{center}
    \includegraphics[keepaspectratio=true,width=180mm]{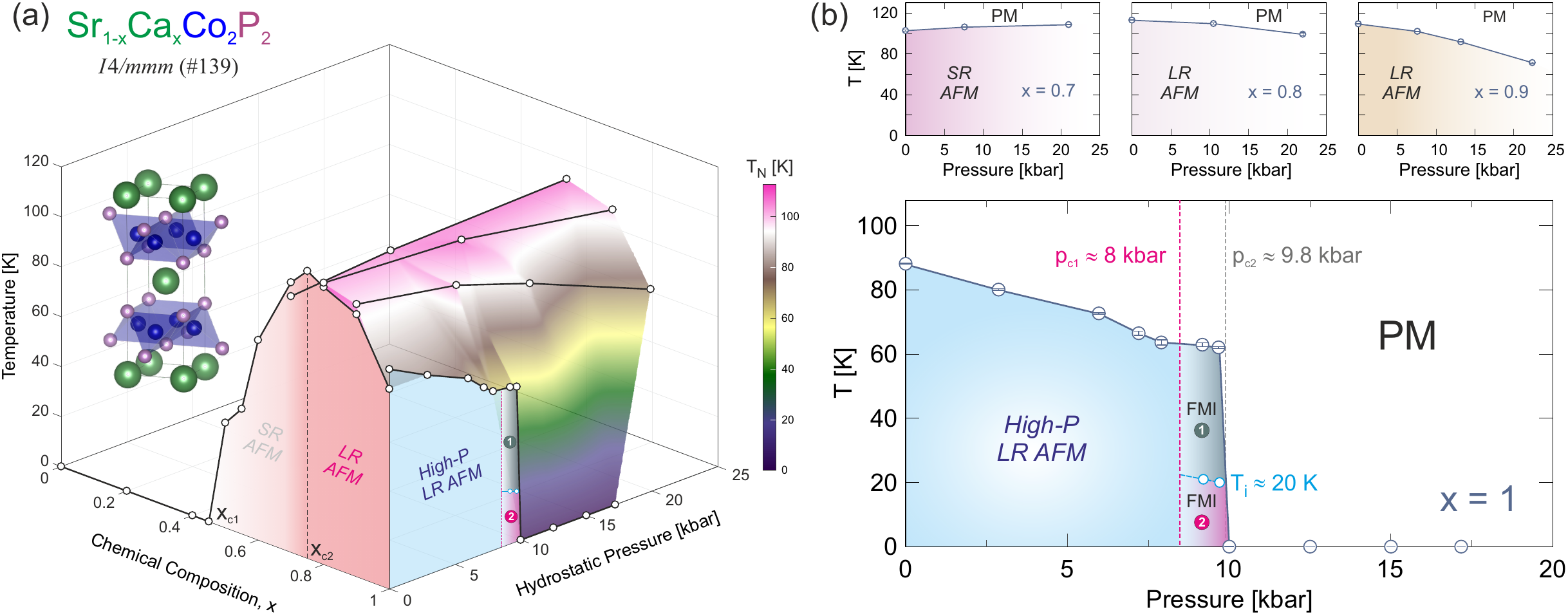}
  \end{center}
  \caption{(a) Phase-diagram ($P/T/x$) for Sr$_{1-x}$Ca$_x$Co$_2$P$_2$. (b) Individual $P/T$ phase diagrams for different chemical compositions ($x$). Phases are labelled: SR = short-range, LR = long-range, AFM = antiferromagnetic, FMI = ferromagnetic islands, PM = paramagnetic. Further, $p_{\rm c1}\approx8$~kbar and $p_{\rm c2}\approx9.8$~kbar are the critical pressures and $x_{\rm c1}\approx0.45$ and $x_{\rm c2}\approx0.75$ critical chemical compositions \cite{Sugiyama2015}. Finally, $T_{\rm i}\approx20$~K is an approximate transition temperature between the FMI-\textcircled{\small{1}} and FMI-\textcircled{\small{2}} phases.
  }
  \label{fig:Phase}
\end{figure*}

Based on the presented pressure dependent results, and from previous ambient pressure study \cite{Sugiyama2015}, a $T/p/x$ phase diagram for Sr$_{1-x}$Ca$_x$Co$_2$P$_2$ can be constructed [Fig.~\ref{fig:Phase}(a)]. In general, the detailed ground state is estimated from the ZF measurements while the temperature boundary is estimated from TF configuration. The transition temperatures, based on TF measurements, are presented in Fig.~\ref{fig:Phase} as a function of pressure. As already pointed out, a strong pressure dependence is observed for the $x=1$ compound. The transition temperature decreases linearly with pressure, until it is completely and suddenly suppressed around $p_{\rm c2}\approx9.8$~kbar. Such critical pressure is fully coherent with the ZF frequency dependence presented above in Fig.~\ref{fig:ZFPara_1}(b). Further, both ZF (Fig.~\ref{fig:ZFPara_1}) and TF (Fig.~\ref{fig:TFPara_1}) data show signatures for additional phases (FMI-\textcircled{\small{1}} and FMI-\textcircled{\small{2}}) appearing already at pressures in the vicinity of $p_{\rm c1}\approx8$~kbar and $T_{\rm i}\approx20$~K. The origin of such phases are further discussed below in Sec.~\ref{sec:discussion}.

A linear like decrease in the transition temperature is also observed for the $x=0.9$ and $x=0.8$ compounds [Fig.~\ref{fig:Phase}(b)]. While the magnetic order is persistent within the current pressure range, most likely, the long range order will be destroyed at higher pressures. More unexpectedly, the pressure dependence of the $x=0.7$ compound is opposite from the other ones. Instead of a decrease in transition temperature, the pressure slightly increases $T_{\rm N}$, suggesting that the magnetic order is stabilized under pressure. Such behaviour is consistent with the results obtained at ambient pressure, where the $T_{\rm N}$ of $x=0.7$ is lower than that of $x=0.75$, at which point a long range magnetic order is stabilised \cite{Sugiyama2015}. In other words, chemical pressure (and hydrostatic pressure) stabilises the magnetic order for $x<0.75$ until a long range order is stabilised. That being said, the ZF time spectra [Fig.~\ref{fig:ZFSpec}(d)] and the fit results show no apparent change with pressure. Therefore, a short range order can be expected up to at least 20~kbar for the $x=0.7$ compound. 


\section{\label{sec:discussion}Discussion}
Given that no significant pressure dependence is observed for $x\neq1$ compounds, we may attribute the high pressure ground state (up to 20 kbar) to be the same as in ambient pressure. Previous ambient pressure $\mu^+$SR study \cite{Sugiyama2015} as a function of $x$ determined the formation of short range AF order for $0.45<x\leq0.75$, which develops into an long range AF order for higher $x$. Such findings is consistent with this study, where distinct oscillations are present for $x>0.7$ compounds, but only a single highly damped oscillation is present for the $x=0.7$ compound. The magnetic order formation was found to be strongly correlated with the distance of Co ions and the adjacent Co$_2$P$_2$ planes ($d_{\rm Co-Co}$). In detail, a decrease in $d_{\rm Co-Co}$ is observed from about $x=0.4$ and decreases linearly up to about $x=0.8$. 

If the magnetic order is truly only dependent on $d_{\rm Co-Co}$, then one would expect the formation of long range order at higher pressures for the $x=0.7$ compound. Instead, only small pressure dependence is observed, even though the $x=0.7$ seems slightly more stable at higher pressures. This would suggest that the exchange mechanism that stabilises the magnetic order does not only dependent on $d_{\rm Co-Co}$. Of course, we should acknowledge the fact that the pressure applied in this study is hydrostatic and not uni-axial, even though chemical pressure can be considered equivalent to hydrostatic pressure. However, it is of course possible that a LRO is in fact stabilised at even higher pressures. Such premise may be confirmed by a pressure study on the $x=0.75$ compound, which is closer to the LRO phase and should yield lower critical pressures. It is noted that while the frequency of the short range ordered phase is small, the highly damped nature does not stem from magnetic inhomogeniousity. Instead, the highly damped nature stems from a broad field distribution due to SRO formation, whereas the value of the frequency itself suggests that the internal field at the muon site is small. 

The $x=1$ compound on the other hand exhibits a strong pressure dependence, where the external pressure destabilizes the magnetic order. Such destabilizing with pressure was also observed for $x=0.9$ and 0.8 compounds, even though it is much weaker. A simple and rough extrapolation would suggest that the long range order is fully suppressed at around 50 and 100~kbar for $x=0.9$ and 0.8, respectively. The much weaker pressure dependence of the $x=0.9$ and 0.8 compounds supports the fact that the magnetic order may be established by exchanges other than simply the inter-plane Co ion interaction. 

The pressure clearly induces a transition in the $x=1$ compound, from magnetically long-range order to a magnetically disordered state. Suppression of magnetic order under applied external pressure is a signature of quantum criticality \cite{Forslund2019, Sugiyama2020}. Of course, for such scenario one would expect that $T_{\rm N}$ would be more smoothly driven towards $T=0$~K, which is not really seen here. That being said, the related compound CrRh$_2$Si$_2$ \cite{Movshovich1996} or the $d$-electron antiferromagnet Cr$_{1-x}$V$_x$~\cite{Lee2004} were both shown to exhibit a second order nature, despite a similar abrupt decrease. Figure~\ref{fig:ZFPara_1}(b) show the order parameter evolution as a function of pressure for the $x=1$ compound. Here, an abrupt decrease of the order parameter is observed between 9.2 and 9.7~kbar. In other words, the suppression of the magnetic order under pressure is not likely to be quantum critical. This is also coherent with the fact that Fermi liquid behaviour is observed through out the pressure range, based on resistivity measurements \cite{Baumbach2014}. Instead, the transition is first order in nature and thus most likely driven by a structural component. Such driver is consistent with muon site re-population mentioned above. In fact, similar compounds have been shown to undergo first order transition under pressure, e.g. EuCo$_2$P$_2$ and SrNi$_2$P$_2$ but then at higher temperatures~\cite{Huhnt1997}.

Since there is no XRD data under pressure, and at low temperature, available, we should acknowledge the fact that a muon site re-population may originate from magneto structural transition and not simply from a crystalline structural transformation. In other words, the missing fraction may not stem from a change of crystalline muon sites but a change in the magnetic environment. Such behaviour would be consistent with the sample undergoing AF-FM transition under pressure above 6~kbar ($p_{\rm c1}\approx8$~kbar). Typically, missing fractions are also more commonly observed for FM structures rather than in AF. Although, this is not a general rule. One could perhaps expect FM correlations to emerge under pressure at lower temperatures given that FM correlations seem to be present at ambient pressure above $T_{\rm N}$ \cite{Imai2014}. Indeed, the related compound LaCo$_2$P$_2$ is an itinerant FM at ambient pressure \cite{Morsen1988, Reehuis1994}. While any pressure dependent measurement is missing for LaCo$_2$P$_2$, a second phase emerged in the itinerant FM UGe$_2$ at lower temperature close to $p_C$~\cite{Pfleiderer2002}, similar to what is observed for the $x=1$ compound. If the title compound undergoes a AF-FM transition under pressure, the discussions about the QCP and its relationship to the order parameters (Fig.~\ref{fig:ZFPara_1}(b)) is not viable given the fact that the order parameter of the main/new phase (the missing fraction) is not accessible. 

Resistivity measurements under pressure revealed a change in its derivative at 8.9~kbar below 50~K~\cite{Baumbach2014}, which was slowly suppressed in temperature with pressure. While the details and the origin remained unsolved, such feature is here clearly revealed to be of magnetic origin. This is based on an additional exponential component manifested in the ZF time spectrum (Fig.~\ref{fig:ZFSpec_1}) and that $A_{\rm TF}$ (Fig.~\ref{fig:TFPara_1}) reveals an anomaly around this pressure ($p_{\rm c1}\approx8$~kbar) and temperature (which in turn is a consequence of such exponential). Since the amount of missing fraction is temperature independent at $p=9.7$~kbar (Fig.~\ref{fig:ZFSpec_1}), the initial faster relaxation at $p=9.7$~kbar should yield some clues regarding the second phase that manifests itself at low temperatures (below $T_{\rm i}\approx20$~K) and above $p_{\rm c1}\approx8$~kbar.

In order to comprehend the second phase present at high pressure and low temperature, it is imperative to unveil microscopic origin of the exponential. Unfortunately, the presence of the missing fraction makes the determination challenging. 
Regardless, lets first point out that there exist mixed fractions at this pressure: a missing fraction, a disordered phase ($A_{\rm PM}$) and a fraction of muons depolarising in a fast exponential manner, that also changes with temperature. With this in mind, we shall propose few scenarios to the origin of the exponential. \textbf{(I)} The missing fraction is resulting from dilute (ferro)magnetic islands (FMI). In this case, part of the muons will be situated in the disordered phase that is relatively close to two separate yet correlated magnetic islands. The inter-island correlation should depolarise the moun ensemble in an exponential manner that is different from the disordered phase. This scenario implies that some islands merges at lower temperatures (below $T_{\rm i}\approx20$~K), or the inter-island correlations are suppressed at lower temperatures for reasons unknown.  
\textbf{(2)} The exponential is a reminiscence of the missing fraction. It might be that oscillations/fluctuations hidden as a missing fraction is "spilled over" to the time window of $\mu^+$SR. This would imply that the correlation times changes as the temperature is lowered. In other words, the missing fraction is somewhat fluctuating at higher temperatures, that becomes more static at lower temperature. Although, the change in fluctuations has to be driven by other means than just simple thermal effects. 
\textbf{(3)} The muon site coordinates might have temperature dependencies at this pressure. Similar to the pressure dependence of the muon site, the temperature dependence of it could result in re-population of different sites. This implies that the low temperature phase and high pressure phase is driven by a structural component or a spin structure reorientation. 

Given the structural and magnetic degree of freedom present in the $x=1$ compound, it may be that several processes discussed above are viable. In order to discern one scenario from another, it is of high interest to perform high pressure XRD, high pressure magnetisation and neutron diffraction. Although, since the magnetic phases seem to form island like structures, detailed experimental study might prove difficult. We wish to stress that the complex temperature dependence of $A_{\rm TF}(T)$ was not observed for the $x\neq1$ compounds. In other words, impure samples or chemical disorder are not likely the underlying reason behind the features observed here. Especially since the pure $x=1$ compounds are in general cleaner than chemically doped samples. Here, we should emphasize that $\mu^+$SR is able to detect magnetic volume fractions in a sample. To conclude, we propose that the intermediate phase emerging above $p_{\rm c1}\approx8$~kbar consists of dilute ferromagnetic islands (FMI) existing within a disordered phase. Further, such FMI undergo an additional transition between a high- (FMI-\textcircled{\small{1}}) and low-temperature (FMI-\textcircled{\small{2}}) state at $T_{\rm i}\approx20$~K. It should also be emphasized that the values of $p_{\rm c1}\approx8$~kbar and $T_{\rm i}\approx20$~K are approximated from a combination of ZF/TF $\mu^+$SR data (with limited number of pressure points) in combination with the resistivity data from Ref.~\onlinecite{Baumbach2014}. The presented numbers should therefore be taken as estimates and further detailed studies are necessary to more accurately determine the phase boundaries. However, $p_{\rm c2}\approx9.8$~kbar could be considered slightly more well defined.





Finally, we wish to discuss nature of the high pressure phase (above $p_{\rm c2}\approx9.8$~kbar). The resistivity above such critical pressure revealed a hump in the data \cite{Baumbach2014}. Such 'hump temperature' was increasing with higher pressures. One suggestion was that it may be related to a broad magnetic transition, like seen in the related compounds CaNi$_{1-x}$Co$_x$P$_2$ \cite{Jia2010},  BaFe$_{1-x}$Cr$_x$As$_2$ \cite{Marty2011} and BaFe$_{1-x}$Mn$_x$As$_2$ \cite{Thaler2011}. However, such conclusion can be readily excluded based on this $\mu^+$SR study. Instead, it might be that strong electron coupling is behind the such resistivity hump, like in Nb$_3$Sn and Nb$_3$Sb \cite{Fisk1976}, which was also one of the suggestion of Ref.~\onlinecite{Baumbach2014}.


\section{\label{sec:conclusions}Conclusions}
The pressure and temperate dependence on the magnetic nature of Sr$_{1-x}$Ca$_x$Co$_2$P$_2$ for $x=1$, 0.9, 0.8 and 0.7 has been investigated by muon spin rotation, relaxation and resonance ($\mu^+$SR). The weak pressure dependencies for the compounds $x\neq1$ suggests that the rich phase diagram of Sr$_{1-x}$Ca$_x$Co$_2$P$_2$ at ambient pressure may not only be due to chemical pressure effects. The $x=1$ compounds on the other hand exhibits strong pressure effects, where the long range magnetic order present at ambient pressure become fully suppressed at $p_{\rm c2}\approx9.8$~kbar. Intriguingly, two additional phase emerges already just below the critical region, occupying the phase space above $p_{\rm c1}\approx8$~kbar and below $p_{\rm c2}\approx9.8$~kbar. Such results are fully coherent with previous single crystal resistivity measurement under pressure \cite{Baumbach2014}. Since $\mu^+$SR is sensitive to magnetic volume fractions, such phase was proposed to be dilute (ferro)magnetic islands (FMI) co-existing within a disordered phase. It is also revealed that such phase consists of a high- (FMI-\textcircled{\small{1}}) and a low-temperature (FMI-\textcircled{\small{2}}) region, respectively, with a phase boundary at $T_{\rm i}\approx20$~K.


\begin{acknowledgments}

This research was supported by the Swedish Research Council (VR) via a Neutron Project Grant (Dnr. 2016-06955) and the Carl Tryggers Foundation for Scientific Research (CTS-18:272). E.N. is funded by the Swedish Foundation for Strategic Research (SSF) within the Swedish national graduate school in neutron scattering (SwedNess). D.A. acknowledges financial support from the Romanian UEFISCDI project PN-III-P4-ID-PCCF-2016-0112, Contract Nr. 6/2018. Y.S. is funded by VR through a Starting Grant (Dnr. 2017-05078) as well as the Chalmers Area of Advance-Materials Science. J.S. acknowledges support from Japan Society for the Promotion Science (JSPS) KAKENHI Grants No. JP18H01863 and No. JP20K21149.
\end{acknowledgments}

\bibliography{Refs} 
\end{document}